\documentstyle[12pt]{article}                                 

\begin{document}

\newcommand {\bea}{\begin{eqnarray}}
\newcommand {\eea}{\end{eqnarray}}
\newcommand {\be}{\begin{equation}}
\newcommand {\ee}{\end{equation}}

\begin{titlepage}

\title{
\begin{flushright}
\begin{small}
hep-th/9703039\\
PUPT-1684 \\
UPT-739-T\\
\end{small}
\end{flushright}
\vspace{1.cm}
Relativistic Brane Scattering}

\author{
Vijay Balasubramanian
\thanks{Research supported in part by  
DOE grant~DE-FG02-91-ER40671.}\\
\small Joseph Henry Laboratories\\
\small Princeton University\\
\small Princeton, NJ 08540 \\
\small e-mail: vijayb@puhep1.princeton.edu
\and
Finn Larsen
\thanks{Research supported in part by 
DOE grant~AC02-76-ERO-3071.}\\
\small David Rittenhouse Laboratories\\
\small University of Pennsylvania\\
\small Philadelphia, PA 19104 \\
\small e-mail: larsen@cvetic.hep.upenn.edu 
}

\date{ }
\maketitle

\begin{abstract}
We calculate relativistic phase-shifts resulting from the large impact
parameter scattering of $0$-branes off $p$-branes within
supergravity. Their full functional dependence on velocity agrees with
that obtained by identifying the $p$-branes with $D$-branes in string
theory. These processes are also described by $0$-brane quantum
mechanics, but only in the non-relativistic limit. We show that an 
improved $0$-brane quantum mechanics based on a Born-Infeld type 
Lagrangian also does not yield the relativistic results.  
Scattering of 0-branes off bound states of arbitrary numbers of 
0-branes and 2-branes is analyzed in detail, and we find agreement 
between supergravity and string theory at large distances to all orders in
velocity.  Our careful treatment of this system, which embodies 
the 11 dimensional kinematics of 2-branes in M(atrix) theory,
makes it evident that control of $1/n$ corrections will be 
necessary in order to understand our relativistic results within
M(atrix) theory.
\end{abstract}

\end{titlepage}
\newpage

\section{Introduction}
The realization that the Ramond-Ramond $p$-brane solutions to the
$N=2$ supergravities in ten dimensions have weak coupling descriptions
in string theory as $D$-branes has substantially sharpened and
extended the understanding of non-perturbative dualities relating
string theories. The equivalence at large distances between $p$-branes
and $D$-branes was originally established by considering static
forces~\cite{polch95a} and subsequently by scattering various probes
off these objects~\cite{bachas,lensing}.  For example, the large
impact parameter scattering of $D$-branes off $D$-branes has been
analyzed in string theory by evaluating the cylinder exchange
diagram~\cite{bachas,lifschytz}.  The resulting amplitudes are valid
even for relativistic velocities, but the analogous processes in
supergravity have only so far been calculated in the non-relativistic
regime. In this paper we extend the classical results and show
that, as expected, the equivalence between $p$-branes and $D$-branes
continues to hold in relativistic processes. Although conceptually this
agreement is a simple kinematical one, it nevertheless involves rather
non-trivial functional dependences on the velocity of the probe.

$D$-branes also serve as probes that sense spacetime at distances smaller 
than the string scale~\cite{shenker,kp,dfs,scales}. The identification
of the new, shorter length scale as the Planck length
of $11$ dimensional supergravity is particularly interesting in view
of the relationship of the string theories to $11$ dimensional supergravity
and its conjectured quantum description as $M$-theory~\cite{ht,witten}.
This has led to the proposal that $M$-theory in the infinite momentum frame
is defined non-perturbatively by M(atrix)-theory, identified as 
the $n\rightarrow\infty$ limit of the $SU(n)$ quantum mechanics 
describing a collection of $n$ $D0$-branes~\cite{bfss}. An important 
test of this proposal is the recovery of the supergravity scattering 
amplitudes at large impact parameter and small velocities from a 
description using only light open strings. However, as we shall show
in detail, this agreement does not immediately extend to the complicated 
relativistic velocity dependence. 

The disagreement we discuss is a direct consequence of the Galilean,
rather than Lorentzian, invariance in the effective $SU(n)$ quantum
mechanics.  A suitable resummation of the open string theory might
lead to a matching with supergravity.  So we consider $0$-brane
scattering in the context of an improved Lagrangian - a
supersymmetrized nonabelian Born-Infeld action that is manifestly
relativistically invariant.  However, we find that this simple remedy
is not sufficient to recover the relativistic expressions for the
brane scattering phase shifts.

If complete, M(atrix)-theory should be able to reproduce the
relativistic velocity dependences discussed in this paper.  At
leading order in velocity, M(atrix)-theory relies on a high degree of
supersymmetry as well as properties of light-cone kinematics to
reproduce supergravity results at large distances. We will find that,
even when these effects are taken into account, an apparent
disagreement at sub-leading orders in velocity persists. In other
words, the correct relativistic kinematics require either a
substantial modification of the M(atrix) Lagrangian or a highly
non-trivial role of interactions. This conclusion is not surprising as
relativistic invariance is known to be subtle in M(atrix)-theory, if
present at all.  However, it is reached here from a new point of view
that may be helpful for the understanding of M(atrix)-theory dynamics.

The paper is organized as follows.  In Sec.~\ref{sec:classical} we
present the relativistic supergravity phase shifts for scattering of
$0$-branes off $p$-branes. In Sec.~\ref{sec:stringtheory} we recall
the full string theory results for the same processes, as well as
their truncations to the lightest open and closed string modes.
Finally, we discuss as yet unsuccessful attempts to reproduce our
relativistic results from M(atrix)-theory by employing, in
Sec.~\ref{sec:NBI}, a supersymmetrized nonabelian Born-Infeld action
and by carrying out, in Sec.~\ref{sec:matrix}, the highly boosted
M(atrix) kinematics.

\section{Semiclassical Results}
\label{sec:classical}
In this section we consider the large impact parameter scattering of
a $0$-brane probe off a $p$-brane target.  The forces between branes
arise from exchange of gravitons, dilatons, and, in the case of
a $0$-brane target, RR photons.  We find the phase shifts for the
scattering by lifting the process to 11 dimensions where it is purely
gravitational: the probe $0$-brane is a plane-fronted wave, and it
simply follows a null-geodesic.  This 11-dimensional interpretation of
the scattering of branes greatly simplifies our calculations.  We will
completely neglect the backreaction of the probe 0-brane on the
background geometry.  The trivial kinematical backreaction could be
accounted for by using the appropriate reduced mass but we will simply
assume that the target is much heavier than the impinging
probe. Another class of corrections corresponds to radiation emitted
during collision. Such processes could be estimated semiclassically
within the supergravity theory, but their reliable calculation is only
possible in the full string theory.  Fortunately radiative corrections
can be neglected even at arbitrarily high impact velocity as long as
the accelerations remain small. This is certainly the case when the
impact parameter is large as it is in the situations that we study.
Another interesting effect also beyond the semiclassical approximation
is inelastic scattering exciting the internal structure of the target.
This too is expected to be highly suppressed for large impact
parameter scattering.  With these limitations we shall find
phase-shifts for 0-brane p-brane scattering that are valid to leading
order in the impact parameter and to all orders in the probe velocity.
The semi-classical phase shifts are derived from supergravity
using the Hamilton-Jacobi formalism described in the next section.

\subsection{The Hamilton-Jacobi method}
\label{sec:hj}
The Hamilton-Jacobi functional $S$ is a classical functional of
particle path in the classical background. It is convenient to
consider it as an ordinary function that depends on a single spacetime
coordinate and a number of conserved quantities. The entire trajectory
is defined by the requirement that it passes through the spacetime
point, and the conserved quantities parametrize the space of such
trajectories. The function $S$ can be interpreted as the phase of 
a semiclassical wave
function and is computed by solving the Hamilton-Jacobi equation:
\begin{equation}
g^{\alpha\beta}{\partial S\over\partial x^\alpha}
{\partial S\over\partial x^\beta}+m^2 = 0~.
\label{eq:hj}
\end{equation}
This is simply the usual kinematical relation between momentum and
mass and it is convenient to make this manifest by introducing
\begin{equation}
p_\alpha = {\partial S\over\partial x^\alpha}~.
\end{equation}
The Hamilton-Jacobi equation can be solved explicitly with conserved
quantities emerging as integration constants. A parametric form of the
trajectory then follows by differentiation with respect to the
conserved quantities~\cite{landaum,mtw}.

We wish to compute $S$, the phase of a semiclassical wave function for the
scattering of 0-branes from p-brane backgrounds.   A p-brane in $10$
dimensions is described by the background~\cite{hs,dkl,lensing}:
\begin{eqnarray}
ds^2_{10} &=& D_p^{-{1\over 2}}( -dt^2 + dx_1^2+ \cdots + dx_p^2 ) +
     D_p^{1\over 2}  ( dx_{p+1}^2 + \cdots + dx_9^2) 
\label{eqn:extmet} \\
e^{-2\phi_{10}} &=& D^{p-3\over 2}_p 
\label{eqn:dilatonansatz}\\
F_{p+2} &=& \partial_\mu D^{-1}_p\: dt \wedge dx_1 \wedge \cdots
dx_p \wedge dx^\mu \\
D_p &=&  1 +  \frac{q_p}{r^{7-p}}~. 
\label{eqn:harmfct}
\end{eqnarray}
This solution can be lifted to $11$ dimensions as:
\begin{equation}
ds^2_{11} = e^{-{2\phi\over 3}} ds^2_{10}
+e^{4\phi\over 3}(dx_{11}-A_\mu dx^\mu)^2
\label{eqn:kk}
\end{equation}
where $A_\mu$ is a Kaluza-Klein gauge field, interpreted as arising
from $0$-brane RR charge in $10$ dimensions. A three-form field $C$ is
induced for $p=2,4$ but we will not write it explicitly because the
$0$-brane probe does not couple to it.  A 0-brane scattering off a
p-brane in 10 dimensions follows a null geodesic in 
the 11 dimensional background (Eq.~\ref{eqn:kk}) with some fixed momentum
along the compact 11th dimension.  The phase of the semiclassical
wave function of the probe 0-brane is therefore given by solving the
Hamilton-Jacobi equation (Eq.~\ref{eq:hj}) with $m^2=0$.

\subsection{0-2}
As a first example of this formalism, we find the
semiclassical phase shift for the scattering of 0-branes from 2-branes.
The 2-brane background lifted to 11 dimensions is:
\begin{equation}
ds^2_{11}= D^{-{2\over 3}}_2(-dt^2+dx^2_1+dx^2_2)
+D^{1\over 3}_2 (dx^2_3+\cdots +dx^2_{11})
\label{eq:2branemet}
\end{equation}
The coordinates are labelled $(t,x_1,x_2\cdots x_9, x_{11})$ and 
$D_2=1+{q_2\over r^5}$ where $r^2=x^2_3+\cdots+x^2_9$.  An elementary
$2$-brane in $M$-theory has identical form but with the harmonic
function $D_2=1+{\tilde{q}_2\over r^6}$. The expression used here is
recovered by averaging over the $11$th dimension as appropriate for
scattering at large impact parameters from a compactified brane.  We
consider only paths that do not depend on coordinates parallel to the
brane, here $x_1$ and $x_2$.

The classical trajectory of the probe 0-brane in $10$ dimensions
remains in one plane, due to angular momentum conservation. One
angular variable named $\theta$ therefore parametrizes the trajectory.
In this coordinate system the Hamilton-Jacobi equation reads:
\begin{equation}
-D_2 ({\partial S\over\partial t})^2+
({\partial S\over\partial r})^2+
{1\over r^2}({\partial S\over\partial\theta})^2 +
({\partial S\over\partial x^{11}})^2 = 0
\end{equation}
The equation can be solved by separation of variables. 
Conserved quantities
\begin{eqnarray}
{\partial S\over\partial t}&=& -E \label{eqn:consE}\\
{\partial S\over\partial x^{11}}&=& p_{11}\label{eqn:consp11} \\
{\partial S\over\partial\theta}&=& J \label{eqn:consJ}
\end{eqnarray}
appear as separation constants. It is then elementary to solve for 
${\partial S\over\partial r}$:
\begin{equation}
S = -Et + J\theta + p_{11}x^{11}+ 
\int^r dr\sqrt{D_2 E^2-p_{11}^2-{J^2\over r^2}}
\label{eqn:S02}
\end{equation}
The classical trajectory is found in parametric form by differentiating 
with respect to $E$, $J$, and $p_{11}$:
\begin{eqnarray}
t &=& \int^r dr~{D_2 E\over \sqrt{D_2 E^2-p_{11}^2-{J^2\over r^2}}} \\
\theta &=&   \int^r dr~{J\over r^2\sqrt{D_2 E^2-p_{11}^2-{J^2\over r^2}}} \\
x^{11} &=& \int^r dr~{p_{11}
\over \sqrt{D_2 E^2-p_{11}^2-{J^2\over r^2}}} 
\end{eqnarray}
The indeterminate lower limits parametrize the arbitrary origin of
each cyclic coordinate. We will not need these explicit expressions
but it is of conceptual importance that the classical motion is 
present in the formalism.

A scattering phase shift, conventionally denoted $2\delta$, is
obtained by subtracting from $S$ of Eq.~\ref{eqn:S02} the phase
accumulated by a $0$-brane with the same conserved charges moving in a
flat background:
\begin{equation}
\delta_{02}=\int^\infty_{r_{\rm min}} dr
\left[\,\sqrt{D_2 E^2-p_{11}^2-{J^2\over r^2}}
-\sqrt{E^2-p_{11}^2-{J^2\over r^2}}\,\right]
\label{eqn:delta02}
\end{equation}
(The factor of $2$ in $2\delta$ was cancelled by the integral from
$\infty$ to $r_{\rm min}$ along the incoming part of the trajectory).  
As written, the lower integration limit
$r_{\rm min}$ in Eq.~\ref{eqn:delta02} is the classical turning point
where the square root vanishes. The notation is formal because $r_{\rm
min}$ generically takes different values for the two terms which refer
to motion in curved and flat backgrounds.  However, we will only use
Eq.~\ref{eqn:delta02} in the regime of the eikonal approximation where
the energies are so large that the scattering angle remains small and
$r_{min}$ can be taken to be the same for both terms.

\subsection{0-4}
Using the same techniques as in the previous section we can derive the
phase shifts for 0-branes scattering off 4-branes.  A $4$-brane in
$10$ dimensions is lifted to $11$ dimensions as:
\begin{equation}
ds^2_{11}=D_4^{-{1\over 3}}
(-dt^2+dx^2_1+\cdots+dx^2_4+dx^2_{11})
+D_4^{2\over 3}
(dx^2_5+\cdots+dx^2_9)
\end{equation}
where $D_4=1+{q_4\over r^3}$. This is a $5$-brane in $M$-theory,
averaged over one parallel dimension. The corresponding 
Hamilton-Jacobi equation is:
\begin{equation}
-D_4 ({\partial S\over\partial t})^2+
({\partial S\over\partial r})^2+
{1\over r^2}({\partial S\over\partial\theta})^2 +
D_4 ({\partial S\over\partial x^{11}})^2 = 0~
\end{equation}
Conserved quantities are introduced as in 
Eqs.~\ref{eqn:consE}-\ref{eqn:consJ} and the Hamilton-Jacobi functional
follows by quadrature, as before. The phase-shift becomes:
\begin{equation}
\delta_{04}=\int^\infty_{r_{\rm min}} dr  
\left[\sqrt{D_4 (E^2-p^2_{11})-{J^2\over r^2}}
-\sqrt{E^2-p_{11}^2-{J^2\over r^2}}\right]
\end{equation}

\subsection{0-0}
The $0$-brane lifts to:
\begin{equation}
ds^2_{11}=-dt^2+dx^2_1+\cdots+dx^2_{11}
+ (D_0-1)(dt+dx_{11})^2
\label{eqn:0brane}
\end{equation}
where $D_0=1+{q_0\over r^7}$. This metric represents a
gravitational wave in $11$ dimensions, averaged over
its longitudinal direction. Alternatively it can be 
interpreted as a Schwarzchild black hole in $10$ dimensions boosted
by an infinite amount with the black hole mass taken to zero as the boost
parameter is taken to infinity. The Hamilton-Jacobi equation
for another $0$-brane propagating in this background becomes:
\begin{equation}
-({\partial S\over\partial t})^2+
({\partial S\over\partial r})^2+
{1\over r^2}({\partial S\over\partial\theta})^2 +
({\partial S\over\partial x^{11}})^2-
(D_0-1)({\partial S\over\partial t}-{\partial S\over\partial x^{11}})^2 
= 0
\end{equation}
Again the conserved quantities are
Eqs.~\ref{eqn:consE}--\ref{eqn:consJ}.  Note that in this case, unlike
the previous ones, the relative sign of $E$ and $p_{11}$
matters. Identical signs give brane-brane ($0-0$) scattering and
opposite ones give brane-anti-brane ($0-\bar{0}$) scattering.  The
phase-shift becomes:
\begin{equation}
\delta_{00}=\int^\infty_{r_{\rm min}} dr 
\left[\sqrt{E^2-p^2_{11}-{J^2\over r^2}+{q_0 \over r^7}
(E\mp p_{11} )^2}
-\sqrt{E^2-p^2_{11}-{J^2\over r^2}}\right]
\end{equation}
Here the upper (lower) sign corresponds to $0-0$ ($0-\bar{0}$).

\subsection{0-6}
The manifestation in $11$ dimensions of a $6$-brane in $10$ dimensions 
is the Kaluza-Klein monopole:
\begin{equation}
ds^2_{11}=(-dt^2+dx^2_1+\cdots+dx^2_6)
+D_6(dx^2_7+dx^2_8+dx^2_9)+D^{-1}_6(dx_{11}+q_6(1-\cos\theta)d\phi )^2
\end{equation}
where $D_6=1+{q_6\over r}$. In this case the scattering is three
dimensional, as we shall explain below. We employ spherical coordinates
with $\theta=0$ in the initial state and we would have $\theta=\pi$
in the final state if there were no scattering. The azimuthal 
angle $\phi$ is $\phi=0$ in the initial state. The Hamilton-Jacobi 
equation is:
\begin{equation}
-({\partial S\over\partial t})^2+
D_6^{-1}[({\partial S\over\partial r})^2+{1\over r^2}
({\partial S\over\partial\theta})^2+
{1\over r^2}
({\partial S\over\partial\phi}-q_6 (1-\cos\theta)
{\partial S\over\partial x^{11}})^2]+
D_6({\partial S\over\partial x^{11}})^2 = 0
\end{equation}
Here $\theta$ is not a cyclic coordinate so the obvious separation
constants eqs.~\ref{eqn:consE}--\ref{eqn:consp11} and
\begin{equation}
{\partial S\over\partial\phi}=p_\phi
\label{eqn:conspphi}
\end{equation}
must be supplemented with
\begin{equation}
J^2= ({\partial S\over\partial\theta})^2+{1\over\sin^2 \theta}
(p_\phi-q_6 p_{11}(1-\cos\theta))^2
\end{equation}
instead of eq.~\ref{eqn:consJ}.
The Hamilton-Jacobi functional reads:
\begin{eqnarray}
S = -Et + p_\phi \phi + p_{11}x^{11}+ 
\int^\theta & d\theta & \sqrt{J^2-{1\over\sin^2 \theta}
(p_\phi-q_6 p_{11}(1-\cos\theta))^2}+ \nonumber \\
+\int^r & dr & \sqrt{D_6 E^2-D^2_6 p_{11}^2-{J^2\over r^2}}
\label{eqn:S06}
\end{eqnarray}
The $0-6$ scattering is more complicated than previous cases because
the orientation of the orbital angular momentum is not
conserved. However, its magnitude is conserved and the total angular
momentum vector is maintained by the electro-magnetic field
responsible for the interaction between the monopole and the charge.
There is still no loss of generality in choosing coordinates such that
$p_\phi=0$ and it is convenient to do so. However, taking derivatives
with respect to $p_\phi$ before setting it to zero, we see that the
trajectory in $\phi$ coordinates is non-trivial and the motion
accordingly non-planar, as expected when the direction of orbital
angular momentum varies in time. Note that $p_\phi$ is the canonical
momentum rather than the kinematical one; so it can vanish even when there
is motion in the $\phi$ direction. The final state has $\phi_f={\pi\over
2}$ and $\cot {\theta_f\over 2}={q_6 p_{11}\over J}$. This can be
shown explicitly by carrying out the relevant integrals.  (Remarkably,
this also follows using nothing but angular momentum conservation.)
These details are not relevant for what follows and mentioned only for
completeness.

The phase-shift is given as:
\begin{eqnarray}
\delta_{06}=
{1\over 2}\int^{\theta_f}_0 &d\theta & \left[\sqrt{J^2-q_6^2 p_{11}^2
\tan^2{\theta\over 2}}-J\right]+
\nonumber \\ +\int^\infty_{r_{\rm min}} &dr & 
\left[\sqrt{D_6 E^2-D_6^2 p_{11}^2-{J^2\over r^2}}
-\sqrt{E^2-p^2_{11}-{J^2\over r^2}}\right]
\end{eqnarray}

\subsection{The Eikonal approximation}
The conserved quantities are related to initial conditions through:
\begin{eqnarray}
E &=& \mu_0 {1\over\sqrt{1-v^2}} \label{eqn:Einit} \\
p_{11}&=&\mu_0 \label{eqn:p11init} \\
p_{10}&=& \mu_0 {v\over\sqrt{1-v^2}} \label{eqn:p10init} \\
{J\over p_{10}} &=& b
\label{eqn:jorb}
\end{eqnarray}
where $\mu_o$ is the probe $0$-brane mass, $v$ is the probe velocity,
and $b$ is the impact parameter. For convenience we also introduced
the auxiliary quantity $p_{10}$, the magnitude of $10$-dimensional
momentum, defined through $E^2=p_{10}^2+p_{11}^2$. Eq.~\ref{eqn:jorb}
equates the angular momentum denoted $J$ with the magnitude of the
orbital angular momentum $\vec{r}\times\vec{p}$. In the $0-6$ case 
a careful distinction was made between particle angular momentum 
and total angular momentum, but conventions were 
chosen so that Eq.~\ref{eqn:jorb} remains valid in this case.

The eikonal approximation is valid when the interaction is a small 
perturbation in the sense that the two square roots in the expressions
for the phase shifts are comparable. Then the first square root can be 
expanded and the lower limit can be taken at $r_{\rm min}=b$. For
example the phase shift Eq.~\ref{eqn:delta02} becomes
\begin{eqnarray}
\delta_{02} &\simeq & {E^2\over 2} \int^\infty_b dr~
{q_2\over r^5}{1\over\sqrt {p_{10}^2-{J^2\over r^2 }}} \\
& = & {q_2 \mu_0 \over 2v\sqrt{1-v^2}}~{1\over b^4}~I_2
\label{eqn:d02}
\end{eqnarray}
where
\begin{equation}
I_p = \int^\infty_0 dx {1\over (x^2+1)^{7-p\over 2}}
= {1\over 2}\sqrt{\pi} {\Gamma({6-p\over 2})\over\Gamma({7-p\over 2})}~.
\label{eq:Ip}
\end{equation}
The substitution $z^2=r^2-{J^2\over p^2_{10}}$ is helpful in
the intermediate step. Analogous calculations in the other cases yield 
\begin{eqnarray}
\delta_{00} &\simeq & {1\over 2} q_0 \mu_0
{(1\mp\sqrt{1-v^2})^2\over v\sqrt{1-v^2}}~{1\over b^6}~I_0~~~(BB/BA) 
\label{eqn:d00} \\
\delta_{04} &\simeq & {1\over 2} q_4 \mu_0
{v\over \sqrt{1-v^2}}~{1\over b^2}~I_4 
\label{eqn:d04}
\\
\delta_{06} &\simeq & {1\over 2} q_6 \mu_0
{2v^2-1\over v\sqrt{1-v^2}}~I_6
\label{eqn:d06}
\end{eqnarray}
In the case of $0-6$ the complications due to angular momentum
exchange are of higher order in $q_6$ and are not captured by the
eikonal approximation. Note that $\delta_{06}$ is in fact divergent
because $I_6$ is logarithmically divergent. The formal expression for
the phase shift is nevertheless of interest as it will also emerge
(with the same caveats) from string theory.  With our conventions a
positive phase shift signals an attractive force. Only the $6$-brane
is repulsive, and then only for small $v$. Note however that the
concept of a force, and that of a potential, is quite subtle in this
context because of the velocity dependent nature of the
interaction. The phase shifts are unambiguous, of course, and a
potential can be formally defined from them.

It is instructive to compare the validity of the eikonal approximation
for different values of $p$.  The requirement is that
we should be able to expand the square root in the phase shift. 
Using the fact that
$(p_{10}^2-J^2/r^2)$ is $O(p_{10}^2)$ throughout the important
region of the phase shift integral we can easily show that
the eikonal expansion can be made when\footnote{A careful examination 
is required to reach this 
conclusion because $p_{10}^2-J^2/r^2=0$ at the turning point.
However, it is sufficient that $p_{10}^2-J^2/r^2$ remains of
$O(p_{10}^2)$ in the bulk of the important region of the integral.}:
\begin{equation}
{q_p\over b^{7-p}}\ll v^\alpha
\label{eqn:apxs}
\end{equation}
Here $\alpha=2$ for ($0-\bar{0}$, $0-2$, $0-6$), $\alpha=0$ in the
$0-4$ case, and $\alpha=-2$ for $0-0$. In the fully relativistic
regime where $v\sim 1$ the distinctions between the various cases
disappear. For very small velocities we see that, for $0-0$, the
eikonal approximation remains valid even for impact parameters much
smaller than the scale set by $q_p$. In string theory this raises the
interesting possibility of a simple approximation scheme, valid well
below the string scale.  The precise conditions obtained in
Eq.~\ref{eqn:apxs} from kinematical reasoning agree with those
observed in~\cite{scales}. There they were understood 
from non-renormalization theorems by noting that,
in the limit of vanishing velocity, the different cases
$p=\bar{0},2,6$, $p=4$, and $p=0$ preserve none, $1\over 4$, and
$1\over 2$ of the supersymmetry, respectively.

\subsection{Charge quantization}
\label{sec:chargequant}
In the string theoretic calculation (reviewed in 
Sec.~\ref{sec:stringtheory}) the velocity dependence
of phase shifts is parametrized by the function:
\begin{equation}
F_p (v) = 
{(8-2p) + 4\sinh^2 \pi\epsilon \mp \delta_{p,0} 8\cosh\pi\epsilon \over
4\sinh \pi\epsilon }
\label{eqn:Ffunct}
\end{equation}
where $ \cosh\pi\epsilon = {1\over\sqrt{1-v^2}}$.  The supergravity
phase shifts can also be written in terms of $F_p (v)$ as:
\begin{equation}
\delta_{0p} = {1\over 2}~{q_p \mu_0\over b^{6-p}} I_p F_p (v)
\end{equation}

Recall that $q_p$ are charges that appear as coefficients in the
harmonic functions $D_p=1+{q_p\over r^{7-p}}$. As shown in
Appendix~\ref{sec:chargederiv} using purely semiclassical reasoning,
the $q_p$ are related to the quantized charge {\it i.e.} the number of
branes $n_p$, through:
\begin{equation}
q_p = {l_s^{7-p}~g\over 2\pi\omega_{6-p}} n_p
\label{eqn:quant}
\end{equation}
The string length is $l_s = 2\pi\sqrt{\alpha^\prime}$ and the volume 
of $S_{6-p}$ is:
\begin{equation}
\omega_{6-p} = {2\pi^{7-p\over 2}\over \Gamma({7-p\over 2})}
\label{eqn:volume}
\end{equation}
We also need the 0-brane mass $\mu_0={2\pi\over l_s g}$. 
In terms of these quantities, the supergravity phase shifts are:
\begin{equation}
2\delta_{0p} = {\Gamma({6-p\over 2})\over 4\pi^{6-p\over 2}}~ 
({l_s\over b})^{6-p}~n_p~F_p (v)~.
\label{eqn:sugraexp}
\end{equation}
In this form the supergravity phase shifts can be compared with results from
string theory.

\section{String Theory}
\label{sec:stringtheory}
The $0-p$ brane scattering phase-shifts have also been calculated
in string theory to the leading order in the genus 
expansion~\cite{bachas,lifschytz}. The result is:
\begin{equation}
2\delta_{0p} = \int_0^\infty {dt\over 2\pi t} 
e^{-{tb^2\over 2\pi\alpha^\prime}} B_pJ_p
\label{eqn:stringexp}
\end{equation}
where
\begin{eqnarray}
B_p &=&f_1^{-(8-p)}(q) f_4^{-p}(q)
{\Theta_1^\prime (0 | it)\over \Theta_1(\epsilon t| it)} \\
J_p &=& {1\over 2}[-f_2^{8-p}(q) f_3^p (q) 
{\Theta_2(\epsilon t| it)\over \Theta_2 (0 | it) }+
f_3^{8-p}(q) f_2^p(q) 
{\Theta_3(\epsilon t| it)\over \Theta_3 (0 | it) }\mp
\delta_{p,0}f_4^8(q)
{\Theta_4(\epsilon t| it)\over \Theta_4 (0 | it) }] \nonumber 
\end{eqnarray}
The functions $f_i$ and $\Theta_i$ are modular functions
in the standard notation~\cite{dnotes2,gsw2} and $q=e^{-\pi t}$. 

\subsection{Closed String Expansion}
\label{sec:closedexp}
At large distances the amplitudes are dominated by the exchange of
massless closed string states with massive states damped at scales
of order $\alpha^\prime$. This regime is dominated by small values of
$t$ in the integral Eq.~\ref{eqn:stringexp}.  The behavior of the
integrand for small $t$ is found by a modular transformation followed
by expansion in $e^{-{\pi\over t}}$.  The result is:
\begin{equation}
B_p J_p \simeq 2^{-{p-4\over 2}} \pi t^{6-p\over 2} F_p (v)
\label{eqn:BJclosed}
\end{equation}
where $F_p$ was introduced in Eq.~\ref{eqn:Ffunct}.
The phase-shift should also be multiplied with $n_p$, the number 
of target branes. Inserting Eq.~\ref{eqn:BJclosed} in 
Eq.~\ref{eqn:stringexp}, and applying the formula
\begin{equation}
\int_0^\infty dt~t^{4-p\over 2}~e^{-{tb^2\over 2\pi\alpha^\prime}}
=\Gamma({6-p\over 2})~({2\pi\alpha^\prime\over b^2})^{6-p\over 2}
\end{equation}
the phase-shift becomes:  
\begin{equation}
2\delta_{0p}  = {1\over 4\pi^{6-p\over 2}} \Gamma({6-p\over 2})~
 ({l_s\over b})^{6-p}~n_p~F_p (v)
\end{equation}
This is the same as the supergravity expression~Eq.~\ref{eqn:sugraexp}.
It has not been checked before that this agreement indeed holds to all
orders in $v$ as it should.

\subsection{Open String Expansion}
At distances much smaller than the string length the interaction is
dominated by open strings stretching between the target and the probe.
The appropriate terms are isolated by expanding the integrand of
Eq.~\ref{eqn:stringexp} for large $t$. We only know the string theory
amplitudes in the eikonal approximation and for $p=\bar{0},2,6$ this
is only valid at distances much larger than the string length; so
there is no regime dominated by open strings and also captured by the
eikonal approximation. Accordingly, for these cases the integration
giving the phase shift Eq.~\ref{eqn:stringexp} diverges at small
$t$ when the large $t$ approximations to the kernels are inserted.
On the other hand, for $p=0$ and $v$ very small, the eikonal
approximation is valid at much smaller distances and there is a regime
where open strings dominate in a controlled way. In fact, this is also
true for $p=4$~\cite{scales}.  Recalling that $\pi\epsilon\simeq v$ for
small $v$ and expanding the integrand of Eq.~\ref{eqn:stringexp} for
large $t$ gives:
\begin{eqnarray}
B_0 J_0 &\simeq & \pi {12+4\cos 2vt-16\cos vt 
\over 2\sin vt}=4\pi {(1-\cos vt )^2 
\over \sin vt} \label{eqn:B0J0} \\
B_4 J_4 &\simeq & 2\pi {1-\cos vt\over\sin vt}
\label{eqn:B4J4}
\end{eqnarray}
Inserting these expressions into Eq.~\ref{eqn:stringexp}, the open
string approximation to the phase shift is obtained. A surprising
feature is noticed directly from Eqs.~\ref{eqn:B0J0}-\ref{eqn:B4J4}:
expanding to the leading order in $v$, the closed string approximation
Eq.~\ref{eqn:BJclosed} is recovered to leading order in $v$ from the
open string string approximation for $p=0,4$. Equivalently, at leading
order in small velocities, the phase shifts at {\it long} distances
agree exactly, including coefficients, with the phase shifts
at {\em short} distances. This means that the short distance, open
string expansion can be used to reproduce the results of the large
distance, closed string expansion.  In the regime of agreement these
calculations also agree with the eikonal approximation to
supergravity. The realization that $0$-branes are able to probe
spacetime at very small distances while, in this manner, capturing low
energy gravity at large distances, has led to the idea that $0$-branes
may have a particularly fundamental role in the final theory.  The
next section studies approaches by which we can try to extend the
phase shift agreements between the open string approximation and
supergravity to beyond the leading order in velocity.

\section{$0$-brane Quantum Mechanics}
\label{sec:NBI}

In the preceding section the open string approximation to the kernels,
Eq.~\ref{eqn:B0J0} and Eq.~\ref{eqn:B4J4}, was found by explicitly
expanding the exact one-loop expression Eq.~\ref{eqn:stringexp}. There
is an alternative perspective on the truncation to open string modes
that leads directly to the same expressions~\cite{scales}. Here one
considers the effective Lagrangian governing the low velocity
interaction of $n$ $0$-branes:
\begin{equation}
{\cal L}= -{1\over 4}~{\rm Tr}F_{\alpha\beta}F^{\alpha\beta}
+\bar{\lambda}\Gamma^\alpha D_\alpha\lambda
\label{eqn:0branelag}
\end{equation}
where 
\begin{eqnarray}
F_{\alpha\beta} &=&
\partial_\alpha A_\beta-\partial_\beta A_\alpha+[A_\alpha ,A_\beta] \\
D_\alpha\lambda &=& (\partial_\alpha +A_\alpha)\lambda
\end{eqnarray}
should be truncated to quantum mechanics ({\it i.e.} spatial
derivatives omitted). The fields are in the fundamental representation
of SU(n). This
Lagrangian represents, in a condensed form, all the amplitudes that
can be derived using only the lightest open strings running between
$n$ 0-branes. For example, $0-0$ scattering with velocity $v$ and
impact parameter $b$ can be described in this formalism by imposing
the VEVs:
\begin{eqnarray}
 \langle X_1
\rangle &=& vt~{1\over 2}\sigma^{3}
\label{eqn:vevvt}\\
\langle X_2 \rangle &=& b~{1\over 2}\sigma^{3}
\label{eqn:vevb}
\end{eqnarray}
on the $SU(2)$ fields $X_\alpha=2\pi\alpha^\prime A_\alpha$.  The only
subtlety that enters the calculation is that gauge fixing and the
accompanying ghost terms must be considered carefully. (Some
intermediate steps are written explicitly in~\cite{mathur1}).  The
leading contributions to phase shifts derive from fluctuations and can
be expressed as determinants arising from the quadratic terms in the
Lagrangian. The heat kernel representation of these determinants
reproduces the integral Eq.~\ref{eqn:stringexp} with kernel
Eq.~\ref{eqn:B0J0}. Specifically, the phase shift at large impact
parameter derived from the 0-brane Lagrangian in
Eq.~\ref{eqn:0branelag} agrees at small velocities with the one
derived from supergravity.

It is tremendously interesting that a theory based on such a simple
Lagrangian, with no manifest appearance of
familiar geometric concepts, nevertheless reproduces supergravity at
large distances, although only at small velocities. A natural
elaboration is to attempt also to account for the fully relativistic
phase shift at long distances. One strategy is to augment the
Lagrangian Eq.~\ref{eqn:0branelag} with additional terms. This is
natural because the role of velocity is played by the electric field
in the world volume theory; so relativistic velocities correspond to
strong fields where the effective interaction for $0$-branes is the
Born-Infeld Lagrangian, rather than its weak field limit in 
Eq.~\ref{eqn:0branelag}. Unfortunately, the complete supersymmetric
non-abelian Born-Infeld Lagrangian is not known.  However, any such
Lagrangian must reduce to the known supersymmetrization of the
Born-Infeld Lagrangian in the abelian limit~\cite{susyBI} and should
have a bosonic part that agrees with the recently derived non-abelian
Born-Infeld Lagrangian that is valid up to certain ordering
ambiguities that are yet to be understood~\cite{tseytlinNBI}.  For the
present preliminary investigation we shall assume such a
Lagrangian: 
\begin{equation}
{\cal L}= -\sqrt{-\rm{det}~M_{\alpha\beta}}
=-\sqrt{-(M_{00}-M_{0i}M^{ij}M_{j0})}\sqrt{\rm{det}~M_{ij}}
\label{eqn:nbi}
\end{equation}
where
\begin{equation}
M_{\alpha\beta}=\eta_{\alpha\beta}+F_{\alpha\beta}-
2\bar{\lambda}\Gamma_\alpha D_\beta\lambda
+\bar{\lambda}\Gamma^\gamma D_\alpha\lambda~
\bar{\lambda}\Gamma_\gamma D_\beta\lambda~.
\end{equation}
Here $\alpha,\beta=0,\cdots,9$ and $i,j=1,\cdots,9$.
Assuming VEVs of
the form Eqs.~\ref{eqn:vevvt}-\ref{eqn:vevb} the quadratic part
of the Lagrangian becomes
\begin{eqnarray}
{\cal L}_{\rm quad}&=&-\sqrt{1-v^2}[1+{1\over 1-v^2}[
-{1\over 2}(F_{0i}^2-\langle F_{0i}\rangle^2)+
{1\over 2}v^2 F^2_{1i}]+{1\over 4}F^2_{ij} + \\
&~&{1\over 1-v^2}[\bar{\lambda}(\Gamma_0 D_0-v^2\Gamma_1 D_1 )\lambda
+v \bar{\lambda}(\Gamma_0 D_1-\Gamma_1 D_0)\lambda ]
-\bar{\lambda}\Gamma_i D_i\lambda ]~.\nonumber
\end{eqnarray}
Taking a gauge fixed version of this quadratic action as a starting 
point, determinants can be evaluated as in the weak field calculation
of~\cite{scales} and a phase shift determined. It is:
\begin{equation}
\delta_{00} = \int_0^\infty {dt\over 2\pi t} 
e^{-{tb^2\over 2\pi\alpha^\prime}}~4\pi 
{(1-\cos {v\over 1-v^2}t )^2\over \sin {v\over 1-v^2}t}
\label{eqn:d00rel}
\end{equation}
This phase shift, derived from a supersymmetrized nonabelian
Born-Infeld action, is exactly the same as the integral
Eq.~\ref{eqn:stringexp} with the open-string, non-relativistic kernel
Eq.~\ref{eqn:B0J0} modified by the replacement:
\begin{equation}
v\rightarrow {v\over{1-v^2}}
\label{eqn:subst}
\end{equation}
It is a check on our procedure and  algebra that the integral 
Eq.~\ref{eqn:d00rel} suffers no divergence at small $t$. At large 
impact parameters the phase-shift becomes:
\begin{equation}
2\delta _{00}\simeq
{1\over 2\pi^3} ( {l_s\over b})^6 ~{1\over 4}({v\over 1-v^2})^3
\end{equation}
Unfortunately, this does not reproduce the relativistic expression
\begin{equation}
2\delta _{00}\simeq
{1\over 2\pi^3} ( {l_s\over b})^6 ~{(1-\sqrt{1-v^2})^2\over v\sqrt{1-v^2}}
\label{eq:local00} 
\end{equation}
even though the proposed corrections indeed contribute terms that
are leading in distance and subleading in velocity. Of course the 
non-abelian Born-Infeld action Eq.~\ref{eqn:nbi} is perhaps not entirely 
accurate and the result is therefore preliminary. On the other hand
the correct relativistic generalization of the open string kernel 
(Eq.~\ref{eqn:B0J0}) must have a very restricted form for the 
integral Eq.~\ref{eqn:d00rel} to be finite and it is difficult
to imagine any prescription that would yield Eq.~\ref{eq:local00}.
Our procedure gives a very natural generalization of the
open string kernel given in Eq.~\ref{eqn:B0J0}.
However, this improvement is not
sufficient to recover the full relativistic velocity dependence
for $0-0$ scattering.

It should be emphasized that we do not know any systematic argument
that an agreement with the relativistic results should be expected
from our calculation that only uses an improved Lagrangian for the
massless open string modes in the 0-brane theory.  It is possible, for
example, that the massive states of theory must be included.  Our
point is simply that the relativistic corrections from the open string
point of view probably involve entirely different and more complicated
physics than the non-relativistic treatment.  This was already
suspected of course, but the present calculation adds a new
perspective.

\section{M(atrix) theory}
\label{sec:matrix}
Many of the string theory dualities derive from the relationship of 
the string theories in 10 dimensions to the conjectural M-theory in 
11 dimensions that reduces to supergravity at large scales.  
It has been recently suggested that M-theory can be 
defined non-perturbatively in the infinite momentum frame as the 
large $n$ limit of the $SU(n)$ matrix model
governing the low energy dynamics of a $n$ 0-branes~\cite{bfss}.  For
the M(atrix) proposal to succeed, it must reproduce the 
supergravity phase shifts for scattering of p-branes derived
in this paper.  The fact that the 0-brane quantum mechanics discussed
in Sec.~\ref{sec:NBI} reproduces supergravity results to leading order
in velocity~\cite{scales}, coupled with the kinematics of the infinite
momentum frame, has enabled the M(atrix) model to successfully
reproduce p-brane scattering to leading order in
velocity~\cite{bfss,ab,mathur1}.  In this section we show explicitly 
how this correspondence fails at higher order in velocity and discuss 
what it would take to improve the matching.

The kinematics of the M(atrix) model imply that the 2-branes of the
theory are bound to a large number of 0-branes.  To study such systems
we consider scattering of 0-branes from a bound state of 0-branes and
2-branes. This process can be analyzed in supergravity and in string
theory. The phase-shifts at large distances are complicated functions
of two variables that, as we shall show, agree in the two
descriptions.  For small transverse velocities and large boosts in the
compact 11th dimension, M(atrix) theory reproduces these results;
indeed this is a remarkable success of the proposal. As we shall
discuss it is not yet known how to calculate the full functions of
velocity that we derive here purely within the framework of
M(atrix)-theory.

\subsection{(2+0) Solutions in Supergravity}
The Ramond-Ramond gauge field that couples to the 0-brane in 10
dimensions is simply the Kaluza-Klein component of the metric in the
compactification of 11 dimensional supergravity on a circle.
Therefore, from the point of view of supergravity, a bound state of a
2-brane with a collection of 0-branes can be constructed by boosting
a 2 brane along the compact 11th dimension.
Boosting the membrane in Eq.~\ref{eq:2branemet} gives the solution:
\begin{eqnarray}
ds_{11}^2 &=&  D_2^{-2/3} (-d\tilde{t}^2 + dx_1^2 + dx_2^2) +
D_2^{1/3} (dx_3^2 + \cdots +d\tilde{x}_{11}^2 )  \label{eq:2+0}\\
C_3 &=&  (D_2^{-1}-1)~d\tilde{t} \wedge dx_1 \wedge dx_2 \\
D_2 &=& 1 + {P \over r^5} \equiv 1 + { (Q/ \cosh^2\beta) \over r^5}
\end{eqnarray}
Here the coordinates $\tilde{t}$ and $\tilde{x}_{11}$ are the
boosted ones:
\begin{eqnarray}
\tilde{t} &=& t\, \cosh\beta + x_{11} \, \sinh\beta \\
\tilde{x}_{11} &=& t \, \sinh\beta + x_{11} \cosh\beta 
\end{eqnarray}
and we have restored the $3$-form gauge field that was not written
explicitly in Eq.~\ref{eq:2branemet}.  When $\beta=0$ the $2$-brane
solution (Eq.~\ref{eq:2branemet}) is recovered.  As $\beta \rightarrow
\infty$ with $Q_0=Q\tanh\beta\approx Q$ fixed, Eq.~\ref{eq:2+0}
reduces to the $0$-brane solution lifted to 11 dimensions 
Eq.~\ref{eqn:0brane}, except that the harmonic function
$D_0=1+{q_0\over r^7}$ has been replaced by $D_0^\prime=1+{Q_0\over
r^5}$ due to the compactification of dimensions parallel to the
$2$-brane. The precise relationship between charge parameters is 
$Q_0\omega_4 l_s^2= q_0 \omega_6$ where $\omega_k$ is the volume of the
unit k-sphere (Eq.~\ref{eqn:volume}). This can be derived by requiring 
that the two configurations carry identical quantized charge 
(as defined in Sec.~\ref{sec:chargequant}) or, equivalently, 
that their total charges agree when calculated using Gauss' law.

The solution is compactified to 10 dimensions using 
Eq.\ref{eqn:kk} (for more details see also~\cite{russo}). This
yields 3-form and 1-form gauge fields:
\begin{eqnarray}
C_{tx_1x_2} &=& {\partial\tilde{t}\over\partial t}~C_{\tilde{t}x_1x_2}
= \cosh\beta \, \left( {1\over D_2} -1 \right)
\stackrel{r \rightarrow \infty}{\longrightarrow} - 
{(Q/\cosh\beta) \over r^5} \label{eq:2charge}\\
A_{t} &=& -{g_{11,t}\over g_{11,11}}
=- {(Q \, \tanh\beta) \over r^5} \left(1 + {Q\over
r^5}\right)^{-1}
\stackrel{r \rightarrow \infty}{\longrightarrow} - 
{(Q \,\tanh\beta) \over r^5} \label{eq:0charge}
\end{eqnarray}
As in previous sections we define physical charges using the
asymptotic behaviour of the potentials. With this convention the
2-brane charge is $q_2 = Q/\cosh\beta$ and the 0-brane charge is $Q_0
= Q\,\tanh\beta = q_2 \, \sinh\beta$.  These formulae can be
understood physically as follows\footnote{A similar discussion appears
in~\cite{mathur1} in the context of brane-antibrane scattering in
M(atrix) theory.}. Due to the compactness of the 11th dimension a
single 2-brane is described as a periodic array of 2-branes, each
carrying a charge $q_2 = P$ as measured in the {\em boosted} frame
($\tilde{t},\tilde{x}_{11}$). In the {\em stationary} frame
($t,x_{11}$) the spacing of the periodic array in the 11th dimension
is Lorentz-contracted by a factor of $1/\gamma = 1/\cosh{\beta}$.  The
density of 2-branes is increased accordingly so that, after
compactification, the apparent 2-brane charge becomes $q_2 = P \,
\cosh\beta = Q / \cosh\beta$ as derived explicitly in
Eq.~\ref{eq:2charge}. Similarly the 0-brane charge can be understood
by writing it as $Q_0 = q_2 w/\sqrt{1-w^2}$ where the velocity $w$ of
the 2-brane in the 11th dimension is related to the boost through
$\cosh\beta={1\over\sqrt{1-w^2}}$. Since the charge and mass of the
2-brane are equal we recognize the 0-brane charge as being simply
equal to the relativistic momentum of the 2-brane in the 11th
dimension.

\subsection{0-(2+0) Scattering in Supergravity}
Having obtained the (2+0) solution in 11 dimensions in
Eq.~\ref{eq:2+0} it is straightforward to apply the Hamilton-Jacobi
method of Sec.~\ref{sec:hj} to analyze scattering of 0-branes
from this background.      The 0-brane trajectory is planar
and the  Hamilton-Jacobi equation becomes:
\begin{eqnarray}
-D_2 \left({\partial s \over \partial t} \right)^2 +
D_2 \left( {\partial S \over \partial x_{11}} \right)^2 +
\left({\partial S \over \partial r} \right)^2
+ {1 \over r^2} \left( {\partial S \over \partial \theta} \right)^2
\nonumber \\
- (D_2 -1 ) \left( {\partial S \over \partial t} \sinh\beta -
{\partial S \over \partial x_{11}} \cosh\beta \right)^2 = 0
\end{eqnarray}
Using the definitions of conserved quantities 
Eqs.~\ref{eqn:consE}-\ref{eqn:consJ}  and the kinematical relations
Eqs.~\ref{eqn:Einit}-\ref{eqn:jorb}, we find the eikonal approximation
to the supergravity phase shifts as in previous sections:
\begin{eqnarray}
\delta_{0,0+2} &\approx& \int_{r_{min}}^\infty
\left( {Q/\cosh^2\beta \over 2r^5} \right) 
\frac{(E\, \sinh\beta - p_{11} \cosh\beta)^2 + (E^2 - p_{11}^2) }
     {\sqrt{ (E^2 - p_{11}^2) - J^2/r^2}}
\\
&=&
\left( {(Q / \cosh^2\beta) \over 2} \right) {1\over b^4} \mu_0 I_2
\frac{v^2 + (\sinh\beta - \cosh\beta \sqrt{1-v^2})^2 }
     {v \sqrt{1-v^2}}
\end{eqnarray}
Here $v$ is the velocity of the probe 0-brane, $b$ is the impact
parameter and $I_2$ is defined in Eq.~\ref{eq:Ip}.  Introducing a
boost parameter $v = \tanh\pi\epsilon$ to describe the 10 dimensional
velocity of the 0-brane, after a little algebra the phase shift for
0-(2+0) scattering can be compactly written as:
\begin{equation}
2\delta_{0,0+2}\approx {\mu_0 I_2 \over b^4}~Q~
{(\tanh\beta - \cosh\pi\epsilon)^2 \over \sinh\pi\epsilon}
\label{eq:raw}
\end{equation}
The construction of the (2+0) solution related the boost 
parameter $\beta$ to the charge parameters
as $\sinh\beta={Q_0\over q_2}$ and we also found that
$Q_0\omega_4 l_s^2= q_0\omega_6$. From these relations and the 
quantization conditions of Sec.~\ref{sec:chargequant} we find that 
the boost in the compact 11th dimension is quantized 
as $\sinh\beta={n_0\over n_2}$.

Putting all of this together we can rewrite Eq.~\ref{eq:raw} in terms
of the number of 2-branes and 0-branes as:
\begin{equation}
2\delta_{0,0+2}\approx \sqrt{n_0^2+n_2^2}~
{1\over 4\pi^2}~({l_s\over b})^4~{\sqrt{1-v^2}\over v}~ 
({1\over\sqrt{1+({n_2\over n_0}})^2}-{1\over\sqrt{1-v^2}})^2
\label{eq:final002}
\end{equation}
This equation gives the phase shift for large impact parameter
to all orders in velocity for scattering 
of a 0-brane off a bound state of $n_0$ 0-branes and $n_2$ 2-branes.
In the 2-brane limit ($n_2 \rightarrow\infty$) and in the 0-brane 
limit ($n_0 \rightarrow\infty$), the phase shift reduces to the 
0-2 and 0-0 results in Eq.~\ref{eqn:sugraexp}. Note that, when inspecting
Eq.~\ref{eqn:sugraexp} in the $0-0$ case, the two compact dimensions 
parallel to the 2-brane must be averaged over, resulting in 
the replacement $({l_s\over b})^6\rightarrow {\pi\over 2}({l_s\over b})^4$.
An important point is that for general $\sinh\beta={n_0\over n_2}$ 
the phase shift Eq.~\ref{eq:final002} cannot be separated unambiguously 
into 0-brane and 2-brane contributions. 

\subsection{0-(2+0) Scattering in String Theory}
We can compare the supergravity phase shift in Eq.~\ref{eq:final002}
with the results from string theory where $n_2$ 2-branes are bound to $n_0$
0-branes by turning on a magnetic flux on the
2-brane~\cite{gilad,ab,mathur1}. When the flux is 
$2\pi\alpha^\prime F_{12}=\cot{\pi \eta}$ and the probe velocity 
is $v=\tanh(\pi\epsilon)$ the cylinder amplitude 
for 0-(2+0) scattering becomes~\cite{gilad,mathur1}:
\begin{equation}
2\delta_{0,0+2} = n_2\int^\infty_0 
{dt\over2\pi t} e^{-tb^2/2\pi\alpha'} (BJ)_{0,0+2} 
\end{equation}
where
\begin{eqnarray}
(BJ)_{0,0+2}
&=&
{1\over 2}~{\Theta_1^\prime (0 | it)\over \Theta_1(\epsilon t| it)} 
~[-\left({f_2\over f_1}\right)^6
~i{\Theta_2(\eta t| it)\over \Theta_1 (\eta t| it) } 
{\Theta_2(\epsilon t| it)\over \Theta_2 (0 | it) }+ \nonumber \\
&+&\left({f_3\over f_1}\right)^6 
~i{\Theta_3(\eta t| it)\over \Theta_1 (\eta t| it) } 
{\Theta_3(\epsilon t| it)\over \Theta_3 (0 | it) }
-\left({f_4\over f_1}\right)^6
~i{\Theta_4(\eta t| it)\over \Theta_1 (\eta t| it) }
{\Theta_4(\epsilon t| it)\over \Theta_4 (0 | it) }]
\label{eqn:BJ002} 
\end{eqnarray}
For large impact parameter $b$ it is appropriate to
isolate contributions from the lightest closed strings
as described in Sec.~\ref{sec:closedexp}. This corresponds to
expansion for small $t$ which gives:
\begin{equation}
(BJ)_{0,0+2}\approx 2\pi t^2~\frac{(\cos\pi\eta - \cosh\pi\epsilon)^2}
{\sinh\pi\epsilon \, \sin\pi\eta}
\label{eq:closedker}
\end{equation}
and the phase shift becomes:
\begin{equation}
2\delta_{0,0+2} \approx ~{1\over 4\pi^2}~({l_s \over b})^4~ 
\frac{(\cos\pi\eta - \cosh\pi\epsilon)^2}
{\sinh\pi\epsilon \, \sin\pi\eta}
\label{eq:stringres}
\end{equation}
The 0-brane charge induced by the flux is
\begin{equation}
n_0={1\over 2\pi}\int F = 
2\pi\alpha^\prime n_2 F_{12} = n_2\cot\pi\eta
\label{eq:0flux}
\end{equation}
Here we have used the fact that in this paper the moduli were set to
unity to avoid cumbersome notation; {\it i.e.} the lengths of the
compact dimensions were chosen to be $l_s=2\pi\sqrt{\alpha^\prime}$
giving a volume of $l_s^2$ for the 2-torus on which the 2-brane is
wrapped.  In the supergravity calculations 0-brane charge arises
from momentum in the 11th dimension and is parametrized by the boost
$\beta$. Using Eq.~\ref{eq:0flux} and the relation
$\sinh\beta=n_0/n_2$ from the previous section we find that
$\cot\pi\eta = \sinh\beta={n_0\over n_2}$.  The physical
interpretation is simply that the momentum density on the 2-brane 
is the same as the 0-brane charge density on the 2-brane. 
Trigonometric identities
and the relation $\cosh\pi\epsilon={1\over\sqrt{1-v^2}}$ now let us
rewrite the string theory phase shift as:
\begin{equation}
2\delta_{0,0+2}\approx \sqrt{n_0^2+n_2^2}~
{1\over 4\pi^2}~({l_s\over b})^4~{\sqrt{1-v^2}\over v}~ 
 ({1\over\sqrt{1+({n_2\over n_0}})^2}-{1\over\sqrt{1-v^2}})^2
\end{equation}
This is exactly the same as the supergravity result
Eq.~\ref{eq:final002}.  Previous comparisons between supergravity and
string theory at large distances often assumed small impact velocity and
large boost in the 11th dimension (i.e. $n_0\gg n_2)$.\footnote{After
this paper was completed we became aware that the
preprint~\cite{gilad} compares the full dependence on $n_0$ and $n_2$,
at leading order in velocity for the closely related process of
2-branes scattering off a (4+2) bound state.  The same paper also
compares the relativistic scattering of 0-branes from a (4+2) bound
state and reports agreement between the velocity dependent potential
computed from string theory and supergravity to all orders in
velocity.} Here we find agreement between the full functional
dependence on both these parameters.

\subsection{0-(2+0) Scattering in M(atrix) Theory}
The bound state between a 2-brane and large numbers of 0-branes is of
particular interest because it is this composite object that appears
in M(atrix)-theory where the fundamental objects - indeed the only
objects - are $0$-branes.  In M(atrix) theory the calculation of 
phase shifts proceeds using the $0$-brane quantum mechanics of 
Sec.~\ref{sec:NBI}. In this formalism an admixture of two-branes 
can be included by introducing fluxes in the large $n_0$ $0$-brane 
theory.  The precise correspondence between the fluxes and the 
charges is~\cite{bfss,fluxes}: 
\begin{equation}
{1\over 2\pi\alpha^\prime}[X_1,X_2]=i\tan\pi\eta~{\bf I}=
i{n_2\over n_0}~{\bf I}
\label{eq:magvev}
\end{equation}
The last equality is identical to 
Eq.~\ref{eq:0flux}; so the parameter $\eta$ introduced here can 
be identified with the one that entered the string theory calculation. 
Note however that, because of the boosted kinematics in M(atrix) theory,
it is assumed {\it a priori} that $\tan\pi\eta\approx\pi\eta\ll 1$ and 
also that the transverse velocity of the probe 0-brane is 
$v=\tanh\pi\epsilon \simeq\pi\epsilon\ll 1$. The phase shift for
scattering a 0-brane off this configuration of M(atrix) theory
is: 
\begin{equation}
2\delta_{0,2+0} = n_2 \int {dt\over 2\pi t}
~e^{-{b^2 t\over 2\pi\alpha^\prime}}
~4\pi \frac{(\cos\pi\eta t-\cosh vt)^2}{(\sinh vt)~(2\sin\pi\eta t)}
\label{eq:matker}
\end{equation}
In fact, this expression has only been derived for $n_2=1$ from the
Matrix model~\cite{ab,mathur1}.  In that case it follows from the
exact solution of a quantum mechanics problem involving numerous
harmonic oscillators. The numerator derives from small shifts in
the energy levels due to the background fields. The VEV in
Eq.~\ref{eqn:vevb} that introduces the impact parameter acts as a mass
term: hence the exponential damping with parameter $b^2$. Moreover,
the magnetic field (Eq.~\ref{eq:magvev}) also formally introduces a mass
term which however is an operator that, after diagonalization and
summation over the resulting ``Landau'' levels, gives the
$2\sin\pi\eta t$ in the denominator.

The scattering of a 0-brane off a bound state of 0-branes with fluxes
can also be analyzed in string theory and is T-dual to the calculation
described in the preceding section.  In the string formalism
Eq.~\ref{eq:matker} arises as an expansion keeping only the lightest
open string modes of Eq.~\ref{eqn:BJ002}.  In fact we actually arrived
at Eq.~\ref{eq:matker} using this method.  Treating configurations
with $n_2>1$ directly within the Matrix model involves subtleties 
regarding the twisting of states of the 0-brane gauge theory 
with flux that we are in the process of studying.

Treating both $\eta$ and $v$ as quantities of order $1/n_0$ we expand 
Eq.~\ref{eq:matker} for large $n_0$ and find
\begin{equation}
2\delta_{0,0+2}\approx n_0~{1\over 4\pi^2}~({l_s\over b})^4~
{1\over 4v}~[({n_2\over n_0})^2+v^2]^2~.
\end{equation}
This agrees with Eq.~\ref{eq:final002}, expanded for large $n_0$; so,
to  leading order in $1/n_0$ M(atrix) theory reproduces the
supergravity results at large distances.  (Note again that the phase
shifts have only been calculated directly for $n_2=1$ in the M(atrix)
approach.) In previous calculations the supergravity side of this
equality was established by either boosting a potential inferred from
the $0-2$ phase shift~\cite{ab}~\footnote{In this approach a
numerical discrepancy was reported.}, or by expanding
results from closed string theory~\cite{mathur1}.  In contrast we
carried out the boost explicitly on the supergravity solution, checked
that it agrees with the string theory result to all orders in
velocity,  and then proceeded to a direct and successful comparison
between supergravity and M(atrix) theory.

The phase shift at leading order in the impact parameter can be
isolated from Eq.~\ref{eq:matker} by expanding to leading order in
$t$. It is evident that the resulting phase shift includes {\em only}
the leading order term in velocity; so it certainly can not match the
relativistic velocity dependence of the supergravity phase shift.  Of
course, the highly boosted kinematics of the M(atrix) model implies
that terms that are subleading in velocity are suppressed by powers of
$1/n_0$ so that the relativistic corrections are indeed small.  On the
other hand it might have been suspected that using the kinematic trick
of first boosting to the infinite momentum frame, calculating
nonrelativistic amplitudes, and then deboosting would permit a
successful calculation of relativistic amplitudes in the stationary
frame.  Our calculation directly in the boosted frame shows that this
is not possible.

The leading long distance supergravity phase shift in
Eq.~\ref{eq:final002} depends in general on the probe velocity and two
parameters - $n_2$ and $n_0$, the number of 2-branes and the number of
0-branes in the bound state.  The rather intricate dependence on these
parameters reflects the kinematics of scattering in 11 dimensions.
Recovering the complete functional form in the M(atrix) formalism, and
not just the leading terms in $v$ and $1/n_0$, will require control
of $1/n_0$ corrections that has not been achieved at present.  It is
also possible that the M(atrix) Lagrangian will have to be improved,
although the obvious improvement via a nonabelian Born-Infeld action
does not seem to be sufficient, as shown in Section~\ref{sec:NBI}. It
is clear that much more thought is necessary to fully recover the
Lorentz invariant structure of supergravity from M(atrix) theory.  Of
course, this was already recognized for other reasons, but the
scattering calculations of this paper add another perspective and
provide a concrete functional form that should eventually be matched.

\section{Conclusion}
In this paper we have developed a technique for computing, within
supergravity, the large impact parameter phase shifts for scattering
of 0-branes from p-branes and bound states of $p$-branes.  Our results
are accurate to the leading order in impact parameter and to all
orders in velocity.  We have shown that phase shifts computed in
string theory for large distance scattering of D-branes agree to all
orders in velocity with our supergravity results.  The phase shifts
for 0-brane scattering can also be reproduced to leading order in
velocity using the quantum mechanics governing a collection of
0-branes.  Since this leading order matching has been so important in
recent attempts to construct a non-perturbative definition of
M-theory, we attempted to extend the matching to relativistic
velocities.  Our attempt involved a natural extension of the 0-brane
effective Lagrangian to situations involving relativistic velocities -
the nonabelian Born-Infeld action.  Although this improved action
contributed terms to the leading large distance phase shift that were
subleading in velocity , they proved insufficient to restore agreement
with supergravity.  We then also compared the scattering of 0-branes from
the 2-branes of SU(n) M(atrix) theory to the analogous processes in
supergravity.  The kinematics of M(atrix) theory implies that the
2-branes of the theory are bound to a large number of 0-branes. Using
the techniques developed in this paper we computed, to all orders in
velocity, the supergravity scattering of 0-branes from 2-branes bound
to a large number of 0-branes.  The expressions fail to match M(atrix)
theory results beyond the leading order in velocity.  This shows that
interactions of order $1/n$ will have to be accounted for more
carefully in the M(atrix) approach in order to successfully reproduce
supergravity.

\vspace{0.2in} 

{\bf Acknowledgments:} We would like to thank C.~Bachas, J.~Gauntlett,
N.~Manton, W.~Taylor, A.~Tseytlin, and especially C.~Callan for
helpful discussions and comments.  After this paper was submitted to
the archives we became aware of~\cite{gilad} where the complete
relativistic velocity dependent potentials are calculated for
scattering of 0-branes from a (4+2) bound state. F.L. would like to
thank the Isaac Newton Institute for hospitality while part of this
work was carried out.  V.B. is supported in part by DOE grant
DE-FG02-91-ER40671.  F.L. is supported in part by DOE grant
AC02-76-ERO-3071.

\appendix
\section{Derivation of Charge Quantization}
\label{sec:chargederiv}
In this Appendix we turn to the derivation of Eq.~\ref{eqn:quant}. The
standard reasoning is to identify the $p$-branes in supergravity with
$D$-branes in string theory and follow world sheet
considerations~\cite{dnotes2}.  In the following we recover the same
result using semiclassical methods in supergravity, using the method
of~\cite{hair2}.  The $p$-branes
eqs.~\ref{eqn:extmet}--\ref{eqn:harmfct} are solutions to the
equations of motion that follow from:
\begin{equation}
L= {1\over 16\pi G_N}\int d^{10}x
\sqrt{-g}[e^{-2(\phi-\phi_\infty)}R + {1\over 2(p+2)!}F_{p+2}^2 ]
\end{equation}
For $p=0$ this Lagrangian follows from that of pure gravity in 
$11$ dimensions, compactified to $10$ dimensions using the 
Kaluza-Klein {\it ansatz} Eq.~\ref{eqn:kk}. By this reasoning it also 
follows from the periodicity of the $11$th dimension that gauge 
functions satisfy the periodic identification 
$\Lambda\equiv \Lambda+g_{\rm st}l_s$. Insisting that, after further
compactification, the $p$-branes for various values of $p$ are related 
by $T$-duality, the periodicity condition can be extended to all values 
of $p$. This simply assumes that the classical $T$-duality of the full 
Lagrangian can be extended to the semiclassical regime. 
Now, variation and integration by parts gives:
\begin{equation}
\delta L ={1\over 16\pi G_N}\int F^{rt\mu_1\cdots\mu_p} dS_8 \int dt 
\end{equation}
with no sum over repeated indices.  The brane is invariant under gauge
transformations so only the boundary term at infinity
contributes. This is the crucial non-trivial property employed.  As
gauge variation we choose a pure gauge that depends only on time with
$\Lambda_{\mu_1\cdots\mu_p}(\infty)-\Lambda_{\mu_1\cdots\mu_p}(-\infty)=
g_{\rm st}l_s$.  The only non-zero component of the field strength is
$F_{rt\mu_1\cdots\mu_p}=\partial_r D_p^{-1}$ so the charge is
normalized as
\begin{equation}
\int F^{rt\mu_1\cdots\mu_p} dS_8 = q_p (7-p)\omega_{8-p} l_s^p
= 2\pi q_p \omega_{6-p} l_s^p~.
\end{equation}
Signs are of no concern in these manipulations and the position of
indices is irrelevant in flat space. Finally, using the relation
between Newton's coupling constant and the string length ${1\over
16\pi G_N}={2\pi\over g^2_{\rm st}l^8_s} $ the variation becomes
\begin{equation}
\delta L = {2\pi\over g^2_{\rm st} l^8_s}~
2\pi q_p \omega_{6-p} l_s^p~g_{\rm st}l_s = {(2\pi)^2\omega_{6-p}q_p\over
g_{\rm st}l_s^{7-p}}~.
\end{equation}
The invariance of wave functions under gauge transformation quantizes
this variation as $2\pi n_p$ and we find:
\begin{equation}
q_p = {l_s^{7-p}~g\over 2\pi\omega_{6-p}} n_p
\ee
as stated in Eq.~\ref{eqn:quant}. In the derivation we included for 
simplicity no moduli but they could easily be restored. This convention
corresponds to compactification on tori with the selfdual radius
$R=\sqrt{\alpha^\prime}$.

\end{document}